%% file: sample-sigconf.tex
\documentclass[sigconf,screen]{acmart}

\AtBeginDocument{%
  }

\makeatother

\copyrightyear{2023} 
\acmYear{2023} 
\setcopyright{rightsretained} 
\acmConference[CIKM '23]{Proceedings of the 32nd ACM International Conference on Information and Knowledge Management}{October 21--25, 2023}{Birmingham, United Kingdom}
\acmBooktitle{Proceedings of the 32nd ACM International Conference on Information and Knowledge Management (CIKM '23), October 21--25, 2023, Birmingham, United Kingdom}
\acmDOI{10.1145/3583780.3615157}
\acmISBN{979-8-4007-0124-5/23/10}

\author{Xiaojie Sun}
\affiliation{
	\institution{CAS Key Lab of Network Data Science and Technology, ICT, CAS}
	\institution{University of Chinese Academy of Sciences}
	\city{Beijing}
	\country{China}
}
\email{sunxiaojie21s@ict.ac.cn}

\author{Keping Bi}
\affiliation{
	\institution{CAS Key Lab of Network Data Science and Technology, ICT, CAS}
	\institution{University of Chinese Academy of Sciences}
	\city{Beijing}
	\country{China}
}
\email{bikeping@ict.ac.cn}

\author{Jiafeng Guo}
\affiliation{
	\institution{CAS Key Lab of Network Data Science and Technology, ICT, CAS}
	\institution{University of Chinese Academy of Sciences}
	\city{Beijing}
	\country{China}
}
\email{guojiafeng@ict.ac.cn}

\author{Xinyu Ma}
\affiliation{
	\institution{CAS Key Lab of Network Data Science and Technology, ICT, CAS}
 \institution{University of Chinese Academy of Sciences}
 \city{Beijing}
 \country{China}
}
\email{xinyuma2016@gmail.com}

\author{Yixing Fan}
\affiliation{
	\institution{CAS Key Lab of Network Data Science and Technology, ICT, CAS}
 \institution{University of Chinese Academy of Sciences}
 \city{Beijing}
 \country{China}
}
\email{fanyixing@ict.ac.cn}

\author{Hongyu Shan}
\author{Qishen Zhang}
\author{Zhongyi Liu}
\affiliation{%
  \institution{Ant Group}
  \city{Beijing}
  \country{China}}
\email{{xinzong, qishen.zqs, zhongyi.lzy}@alibaba-inc.com}

\usepackage{acronym}

\usepackage{tablefootnote}

\usepackage{bm}
\usepackage{caption}
\usepackage{graphicx}
\usepackage{subfigure}
\usepackage{amsmath}
\usepackage{booktabs}
\usepackage{threeparttable}
\usepackage{multirow}
\usepackage{enumitem}
\usepackage{xcolor}
\usepackage{mathtools}
\usepackage{marvosym}
\usepackage[ruled]{algorithm2e}
\usepackage{tablefootnote}
\usepackage{soul, color, xcolor}

\newcommand{\alipay}{{Alipay}\xspace}
\newcommand{\madr}{\textsc{MADRAL}\xspace}
\newcommand{\our}{\textsc{ATTEMPT}\xspace}

\begin{document}

\title{Pre-training with Aspect-Content Text Mutual Prediction for Multi-Aspect Dense Retrieval}


\renewcommand{\shortauthors}{Trovato and Tobin, et al.}

\begin{abstract}


Grounded on pre-trained language models (PLMs), dense retrieval has been studied extensively on plain text. In contrast, there has been little research on retrieving data with multiple aspects using dense models. In the scenarios such as product search, the aspect information plays an essential role in relevance matching, e.g., category: \textit{Electronics}, \textit{Computers}, and \textit{Pet Supplies}. A common way of leveraging aspect information for multi-aspect retrieval is to introduce an auxiliary classification objective, i.e., using item contents to predict the annotated value IDs of item aspects. However, by learning the value embeddings from scratch, this approach may not capture the various semantic similarities between the values sufficiently. To address this limitation, we leverage the aspect information as text strings rather than class IDs during pre-training so that their semantic similarities can be naturally captured in the PLMs. To facilitate effective retrieval with the aspect strings, we propose mutual prediction objectives between the text of the item aspect and content. In this way, our model makes more sufficient use of aspect information than conducting undifferentiated masked language modeling (MLM) on the concatenated text of aspects and content. Extensive experiments on two real-world datasets (product and mini-program search) show that our approach can outperform competitive baselines both treating aspect values as classes and conducting the same MLM for aspect and content strings. Code and related dataset will be available at the URL \footnote{https://github.com/sunxiaojie99/ATTEMPT}.
\end{abstract}





\maketitle


\vspace{-2mm}
\section{Introduction}
\label{intro}
Dense retrieval models \cite{bibert-1,bibert-2,bibert-3,bibert-4,bibert-5,bibert-6} have achieved compelling performance with pre-trained language models (PLMs) \cite{bert,ernie} as the backbone. 
Most studies on dense retrieval focus on unstructured data consisting of plain text, while little attention has been paid to structured item retrieval such as product and people search. In these scenarios, additional aspect information beyond the query or item content is critical for relevance matching, such as brand-\textit{nike}, 
affiliation-\textit{Stanford}. However, little work has explored how to use them effectively in dense retrieval models.

A typical way of leveraging aspect information for multi-aspect retrieval is to refine the item representations with an auxiliary aspect prediction objective \cite{madr}. 
Specifically, for each aspect of an item, the item content is used to predict its annotated value IDs during training. 
This approach has two major disadvantages: 
1) It considers the values of an aspect as isolated classes and learns the embeddings of value IDs from scratch, ignoring their semantic relations.
For example, among the category values, "Hunting \& Fishing" is more related to "Sports \& Outdoors" while unrelated to "Pet Supplies". However, such semantic relations may not be captured sufficiently if we treat them as independent classes. 
2) It does not use query/item aspects such as category, brand, color, etc. during test time, which limits the potential retrieval gains. 
Although it may be costly to obtain query aspects during online service, item aspects can be extracted offline and it is easy to also use them during inference if they are already used in training. 

In this paper, we propose a method of pre-training with Aspect-contenT TExt Mutual PredicTion (\our) to address the above limitations.  
Specifically, \our leverages aspect values as text strings and concatenates them with the content using leading indicator tokens in between. For more effective retrieval, rather than simply conducting undifferentiated MLM on the concatenated aspect and content text, we specifically design an aspect-content mutual prediction objective. It keeps the entire aspect/content tokens and predicts the masked ones in the content/aspects. Also, to suit the scenario where the overhead of obtaining the query aspects online is high, we set the query aspect text to empty during inference. Our method has several advantages over the common approach: 
1) In \our, the text of an aspect value reuses the token embeddings from the powerful PLMs so the semantic relations between values can be naturally captured. 
2) Being concatenated with the content, the item aspects can also take effect for relevance matching during test time.
3) The aspect-content mutual prediction objective promotes sufficient interactions between the aspect and content at the token level, producing better item representations for retrieval, which is confirmed by extensive experimental results. 


As far as we know, there are no suitable large-scale public datasets for multi-aspect retrieval. We construct such a dataset by crawling the item categories
from their pages to complement the aspects
in the Amazon ESCI dataset \cite{amazon-data}. Our experiments on this refined dataset and a real-world commercial mini-program dataset show that \our can significantly outperform the competitive baselines both predicting the classes of aspect values and conducting the same MLM for aspect and content strings. 
\vspace*{-3mm}

\input{related}
\input{method}
\input{exp-settings}
\input{exp-results}

\bibliographystyle{ACM-Reference-Format}
\bibliography{paper}

\end{document}

%% file: related.tex
\section{RELATED WORK}


There are three threads of work related to our study. 
(1) \textbf{Multi-aspect Retrieval.} 
Some work has exploited multi-aspect information to rank products or entities before PLMs appear \cite{rerank-explain, rerank-explain2, entity-aspect}.
In the era of PLM~\cite{ptm-Fan}, there has been limited research on multi-aspect retrieval until \citet{madr} first attempts to do so.  
They learn aspect embeddings by predicting their value IDs with item contents and fuse them to yield an item embedding. 
Later, \citet{agree} proposed a fine-tuning method that uses the local aspect-level matching signals to enhance the global query-item embedding matching.
(2) \textbf{Multi-field Retrieval.} 
How to effectively leverage multiple fields (e.g., title, body, etc.) in a document has been a long-standing research topic. The most famous method is BM25F \cite{bm25-fields}. Methods leveraging multi-fields have also been proposed before and after PLMs appeared \cite{rerank-mcn,rerank-phc,rerank-NRM,rerank-structed}.
The multi-fields are unstructured text in nature and the essential issue is how to weigh them differently during matching. Aspects, unlike fields, usually have a fixed value set which is much smaller than the space of field text. Thus, their core challenges are different.
(3) \textbf{Pre-trained Models for Dense Retrieval.}
Many studies have explored promoting the capabilities of PLMs for dense retrieval including introducing extra training objectives\cite{costa,pretrain-ict,ORQA,seed,cikm-Ma}, special masking schemes \cite{RetroMAE}, and model architecture changes \cite{condenser}, etc. Our method is grounded on the basic dual BERT encoders \cite{bert}.

%% file: method.tex
\vspace*{-2mm}
\section{METHODOLOGY}
\subsection{Preliminary}
For a query q or a candidate item $i$ 
, we represent 
the content text (e.g., query string, title, description) as $t_{c}$ and the aspect text (e.g., values for brand, color, and category) as $t_{a}$. 
Assuming $q$ or $i$ has $k$ aspects, $t_{a}$ is further denoted as $t_{a_1}, ..., t_{a_k}$. 
For each aspect $a_j$ ($1\le j \le k$), it has a finite vocabulary of aspect values, denoted as $V_{a_j}$. 
Previous work \cite{madr} incorporates aspect information by predicting the IDs corresponding to the annotated values of each aspect $a_j$ within the space $V_{a_j}$. 
In contrast, we propose to pre-train the encoder by conducting mutual prediction between text $t_a$ and $t_c$.

\vspace*{-4mm}
\subsection{\our}\label{sec:al}
\begin{figure}
    \centering
    \setlength{\abovecaptionskip}{0.1cm}
    \includegraphics[scale=0.32]{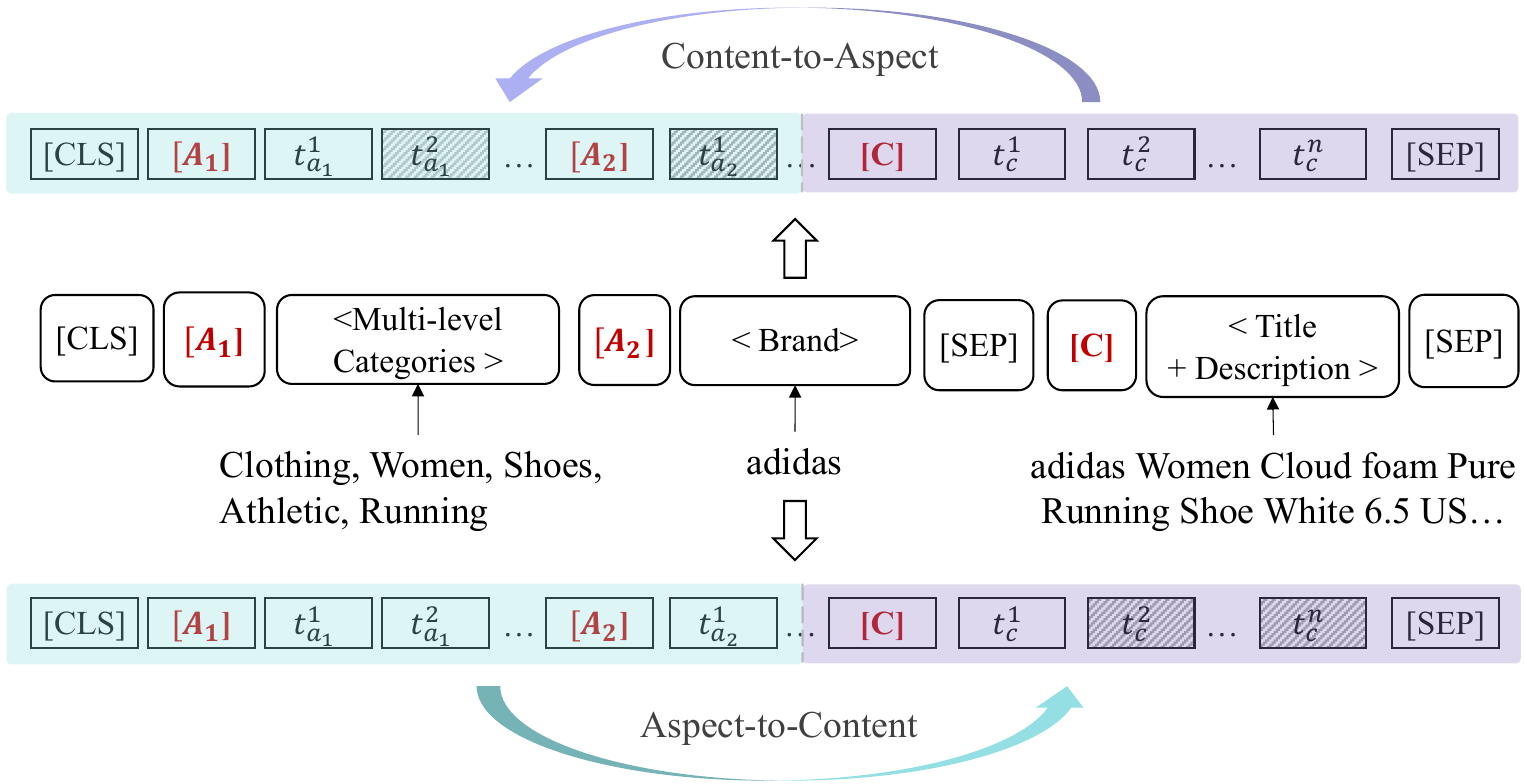}
    \caption{The mutual prediction MLM in \our. The aspect and content texts are colored green and purple.}
    \label{fig:input}
\end{figure}

To model the semantic relationship between various values of an aspect naturally, we treat the aspect values as text strings and concatenate them with the content text. 
For sufficient capture of the interactions between item aspects and contents, we introduce mutual prediction objectives as illustrated in Figure \ref{fig:input}.


\noindent%
\textbf{Encoder Input.} 
To indicate different types of text segments, we prepend an indicator token $[A_j]$ ($1\le j \le k$) and $[C]$ to the aspect text $t_{a_j}$ and the original content $t_c$,
e.g., an encoder input is
$[A_1] t_{a_1} [A_2] t_{a_2} [A_3] t_{a_3} [SEP] [C] t_{c} [SEP]$.
When a query/item does not have certain aspect information, the corresponding aspect text will be empty. In this case, the indicator tokens could still learn some implicit representations of the query/item content. 
Note that during relevance matching, we always keep the query aspect text empty to suit the practical retrieval scenarios where the overhead of obtaining query aspects is high and also avoid potential semantic drift. 
Table \ref{tbl:qd} will show that the query-side indicator tokens ($[A_j]$ $(1\leq j\leq k)$, $[C]$) alone learned during pre-training are beneficial for retrieval.  
Since the other parts of \our are exactly the same between $q$ and $i$, we take $i$ as an example for illustration. 

\noindent%
\textbf{Content Masked Language Modeling (MLM).} 
To capture the interactions between the content tokens without any auxiliary information, 
\our conducts MLM on the item content. It randomly masks tokens in the content text and predicts the masked tokens with the context-dependent representations encoded by Transformer layers \cite{bert}. The corresponding loss function is: 
\begin{equation}\label{mlm2}
\small
\setlength{\abovedisplayskip}{1pt}
\setlength{\belowdisplayskip}{1pt}
\begin{aligned}
\mathcal{L}_{MLM}(\bm{\hat{t_c}})=-\sum_{w \in m(\bm{\hat{t_c}})}logP(w | \bm{\hat{t_c}}_{\backslash m(\bm{\hat{t_c}})}),
\end{aligned}
\end{equation}
where $\bm{\hat{t_c}}$ denotes the text produced by randomly masking some tokens in the text $\bm{t_c}$,
$m(\bm{\hat{t_c}})$ denotes the masked tokens, and $\bm{\hat{t_c}}_{\backslash m(\bm{\hat{t_c}})}$ denotes the remaining tokens in $\bm{\hat{t_c}}$.

\noindent%
\textbf{Aspect-to-Content MLM Prediction.} 
We take the entire aspect text as context when predicting the masked tokens in the content text. 
Under this context, the prediction of masked content tokens has extra evidence for consideration and can act differently than content MLM alone. The aspect-to-content (a2c) loss $\mathcal{L}_{a2c}$ is:
\begin{equation}\label{lossa}
\small
\setlength{\abovedisplayskip}{1pt}
\setlength{\belowdisplayskip}{1pt}
\begin{aligned}
\mathcal{L}_{a2c}(\bm{t_a} \oplus \bm{\hat{t_c}})=-\sum_{w \in m(\bm{\hat{t_c}})}logP(w | \bm{t_a} \oplus \bm{\hat{t_c}}_{\backslash m(\bm{\hat{t_c}})}),
\end{aligned}
\end{equation}
where $\oplus$ means concatenation. In particular, the leading tokens [$A_j$] ($1\le j \le k$) and [$C$] in the input will not be masked.

\noindent%
\textbf{Content-to-Aspect MLM Prediction.}
The idea of content-to-aspect prediction is similar to the aspect classification in \cite{madr}, both of which use the original content to predict the aspects. However, \our predicts the masked words in the aspect text rather than the value classes (IDs), which encodes the aspect information in a softer manner. Specifically, the content-to-aspect (c2a) loss is:

\begin{equation}\label{lossb}
\small
\begin{aligned}
\setlength{\abovedisplayskip}{0pt}
\setlength{\belowdisplayskip}{0pt}
\mathcal{L}_{c2a}(\bm{\hat{t_a}} \oplus \bm{t_c})=-\sum_{w \in m(\bm{\hat{t_a}})}logP(w | \bm{\hat{t_a}}_{\backslash m(\bm{\hat{t_a}})} \oplus \bm{t_c}).
\end{aligned}
\vspace{-1mm}
\end{equation}

\noindent%
\textbf{Overall Learning Objective. } 
By introducing $\mathcal{L}_{a2c}$ and $\mathcal{L}_{c2a}$, \our can incorporate the aspect information into the item representation sufficiently through bidirectional interactions.
In summary, our overall pre-training objective is:
\begin{equation}\label{final}
\small
\begin{aligned}
\mathcal{L}_{overall} = \mathcal{L}_{MLM}(\bm{\hat{t_c}}) + \lambda(\mathcal{L}_{a2c}(\bm{t_a} \oplus \bm{\hat{t_c}}) + \mathcal{L}_{c2a}(\bm{\hat{t_a}} \oplus \bm{t_c})),
\end{aligned}
\end{equation}
\vspace{-2mm}
where $\lambda$ is a hyper-parameter.

%% file: exp-settings.tex
\vspace*{-1mm}
\section{EXPERIMENTAL SETUP}\label{sec:exp_setting}
\vspace*{-1mm}
\subsection{Datasets}
\label{sec:dataset}
We conduct model comparisons on two real-world datasets:

\noindent%
\textbf{Multi-Aspect Amazon ESCI Dataset (MA-Amazon).} 
Amazon ESCI Product Search \cite{amazon-data} originally has multilingual real-world queries, product information such as brand, color, title, description, etc., and 4-level relevance labels: \textit{Exact}, \textit{Substitute}, \textit{Complement}, and \textit{Irrelevant}.
We only use the English part and enrich the dataset by collecting multi-level product categories from the item pages. 
We merge all the items and get a corpus of 482K unique items, which is used for pre-training. For fine-tuning, we divide the original training set into training and validation sets by queries, and keep the test set, yielding 17K, 3.5K, and 8.9K queries respectively.
As in \cite{amazon-data}, we treat \textit{Exact} as relevant and the other labels as irrelevant during training and for recall calculation.
MA-Amazon only has item aspect information, and the coverage of brand, color, and category of levels 1-2-3-4 are 94\%, 67\%, and 87\%-87\%-85\%-71\%, respectively.


\noindent%
\textbf{\alipay Search Dataset.}  
\alipay is a mini-program (app-like service) search dataset with binary manual relevance annotations.
The pre-training query/item corpus has 1.3M/1.8M distinct queries/items with aspect information i.e., \textit{brand} (44\%/0.6\% coverage on query/item) and \textit{three-level categories} (91\%-90\%-56\%/90\%-90\%-62\% coverage for category 1-2-3 of query/item). 
The fine-tuning dataset consists of 60K/3.3K/3.3K unique queries in the training/validation/test set.
Note that the queries for validation and testing do not appear in the pre-training query corpus.
\vspace*{-2mm}
\subsection{Baselines}
We compare \our with the following pre-training methods 
(\textbf{-C} means that the input takes the same concatenation strategy for aspect and content text as \our):
(1) \textbf{BIBERT} \cite{bibert-pretrain,sent-bert}:
BIBERT, the backbone of \our, is a prevalent dense retrieval method for plain text. 
It employs MLM \cite{bert} to pre-train the encoder using the content text of query/item.
(2) \textbf{Condenser} \cite{condenser}: 
It adds a short circuit between the tokens except \texttt{CLS} of the lower layer and the higher layer of BERT \cite{bert} to enhance the final \texttt{CLS} representation.
(3) \textbf{BIBERT-C}:
It only differs from BIBERT in the encoder input. It uses the aspect text in the same way as \our during pre-training and fine-tuning. 
(4) \textbf{BIBERT-C(A)}:
It refines BIBERT-C by assigning a higher mask ratio specifically for the aspect text, which is consistent with \our.
(5) \textbf{MTBERT [-C]} \cite{madr}:
It conducts $k$ additional aspect classification tasks on the \texttt{CLS} during pre-training.
(6) \textbf{\madr [-C]} \cite{madr}: 
It initiates extra multiple aspect embeddings and learns them by predicting the value classes of each aspect and fuses them to produce the final item representation.
\vspace*{-2mm}
\begin{table}
\small
\renewcommand{\arraystretch}{0.4}
\setlength{\abovecaptionskip}{0pt}
\setlength\tabcolsep{3pt} 
\caption{Overall performance. 
The best results are in bold. 
$\dag$ indicates significant differences between \our and the best baselines in the first/second/third group.
}
  \label{main_exp}
\begin{threeparttable}
  \begin{tabular}{lcccccc}
    \toprule
     \multirow{2}{*}{Method} & \multicolumn{3}{c}{MA-Amazon}&\multicolumn{3}{c}{\alipay} \\
     \cmidrule(lr){2-4} \cmidrule(lr){5-7}
      &r@100&r@500 & ndcg@50 &r@100&r@500  & ndcg@50 \\
    \midrule
    BIBERT& 0.6075 & 0.7795 & 0.3929 & 0.4464 & 0.6284 & 0.2033 \\
    Condenser& \underline{0.6091}$^{\dag}$&  \underline{0.7801}$^{\dag}$& \underline{0.3960}$^{\dag}$& \underline{0.4520}$^{\dag}$ & \underline{0.6423}$^{\dag}$ & \underline{0.2072}$^{\dag}$ \\
    \midrule
    MTBERT& \underline{0.6137}$^{\dag}$ & \underline{0.7852}$^{\dag}$ & \underline{0.3969}$^{\dag}$ &  0.4498 & 0.6280 & 0.2064   \\
    \madr& 0.6088 & 0.7815 & 0.3950 & \underline{0.4506}$^{\dag}$ & \underline{0.6383}$^{\dag}$ & \underline{0.2057}$^{\dag}$  \\
    \midrule
    BIBERT-C& 0.6137 & 0.7814 & 0.4005&  0.4517 & 0.6291 & \underline{0.2103}  \\
    BIBERT-C(A)& 0.6137 & 0.7841  & 0.4019 & \underline{0.4611} & \underline{0.6432}$^{\dag}$ & 0.2091  \\
    MTBERT-C& 0.6142 & 0.7839 & 0.3997&  0.4391 & 0.6189 & 0.2026   \\
    \madr-C&  \underline{0.6169}$^{\dag}$ &  \underline{0.7850}$^{\dag}$ &  \underline{0.4041}$^{\dag}$ &0.4376  &0.6141  & 0.2044 \\
    \midrule
    \our& \textbf{0.6233}& \textbf{0.7924} & \textbf{0.4097} & \textbf{0.4667} & \textbf{0.6592} & \textbf{0.2113}   \\
    \bottomrule
  \end{tabular}
  \end{threeparttable}
\end{table}

\vspace*{-2mm}
\subsection{Implementation and Evaluation Details}

We implemented \our and all the baselines by ourselves. 
For all the methods, the encoder is shared for both queries and items.

\noindent%
\textbf{Pre-training.}
The maximum token length is 156, 
The learning rate and epoch for the MA-Amazon/\alipay dataset are set to 1e-4/5e-5 and 20/10, respectively. 
We initialize the BERT parameters with Google's public checkpoint and use Adam optimizer with a linear warm-up.  
For all -C baselines and \our, the mask ratios are set to 0.15/0.3 for item/query content to account for the shorter query length.
They all have the same mask ratio between aspect and content text except for BIBERT-C(A) and \our, where the mask ratio for aspect text is 0.6. $
\lambda$ in Eq.\ref{final} is set as 1.0. 
We fine-tune the pre-trained model checkpoints every two epochs and select the best one on the validation dataset. 

\noindent%
\textbf{Fine-tuning.}
On both datasets, all models are trained for 20 epochs with the Tevatron toolkit\cite{tevatron}. 
We use a learning rate of 5e-6 and a batch size of 64.
All methods are trained with softmax cross entropy loss with in-batch negatives and one hard negative.  
Note that we have not used auxiliary classification objective for MTBERT and \madr since no significant improvements are achieved.

\noindent%
\textbf{Metrics.}
We report recall@100, recall@500 and ndcg@50.
When calculating ndcg on MA-Amazon, following \cite{amazon-data}, we set the gains of E, S, C, and I to 1.0, 0.1, 0.01, and 0.0, respectively.
We perform two-tailed t-tests (p-value $\le$ 0.05) to see significant differences. 


%% file: exp-results.tex

\vspace*{-2mm}
\section{EXPERIMENTAL RESULTS}

\vspace*{-1mm}
\subsection{Main Results}
The overall performance is shown in Table \ref{main_exp}.
We have the following observations: 
(1) Generally, methods using aspect information outperform those that don't, confirming the importance of aspects in relevance matching. Notably, MADRAL performs worse than MTBERT on MA-Amazon, possibly due to insufficient pre-training data to learn the aspect embeddings from scratch sufficiently.
(2) Models treating aspect information as text strings (BIBERT-C/-C(A)) surpass those considering aspect values as discrete classes (MTBERT and \madr). 
When the aspect text has a larger mask ratio than the content (in BIBERT-C(A)), the retrieval performance will be boosted. This shows that the aspect text should be taken special care to encourage sufficient learning. 
(3) When aspect text concatenation is incorporated with methods using aspect values for classification (MTBERT and \madr), the retrieval performance does not always become better. 
This could be because the input aspect text becomes the shortcut for the models to predict its corresponding class ID. When the pre-training data is large (e.g., on \alipay), such relation is more likely grasped by models, deterring the learning of beneficial interactions. 
(4) More powerful pre-training method (Condenser) sometimes perform better than methods using aspects (MTBERT and \madr on \alipay). Note that the benefit from the advanced pre-training techniques is orthogonal to the aspect information and they can be combined for even better performance. We leave the study of this in future work. 
(5) Overall, our \our achieves the best performance on both datasets, showing the efficacy of its pre-training objective specifically proposed for the concatenated text of aspect and content.  
\vspace*{-3mm}
\subsection{Further Analysis}
\label{sec:futher}
We also probe \our from various perspectives to verify its effectiveness.
For reproducibility, our analysis is based on MA-Amazon.  
The only exception is the ablation study of query/item aspects since only \alipay has both of them. 

\noindent%
\textbf{Ablation Study of Aspects.}
We study the effects of various aspects in \our (brand, color, and category from level 1 to 4) in Table \ref{tbl:ab-loss}. 
We find that: 
(1) When use each aspect alone, only the category information enhances model performance.
This might because brand and color are often included in the item content already while the category is extra meta information. 
The observation that category matters the most is consistent with \cite{madr}. 
(2) Combining all aspects outperforms using category only, indicating that brand and color may take a better effect when interacting with the category.
(3) More levels of category information will lead to better performance except that three and four levels have similar results. 
While adding more category levels provides richer information, the reduced coverage (refer to Section \ref{sec:dataset}) might limit the benefits.

\begin{table}[t]
\small
\renewcommand{\arraystretch}{0.4}
\setlength{\abovecaptionskip}{0pt}
  \setlength\tabcolsep{11pt} 
  \caption{Study of various component choices on MA-Amazon. $\dag$ indicates significant improvements over BIBERT.}
  \label{tbl:ab-loss}
  \begin{tabular}{llll}
    \toprule
     & r@100 & r@500 & ndcg@50 \\
    \midrule
    BIBERT & 0.6075 & 0.7795 & 0.3929 \\
    \our & \textbf{0.6233}$^\dag$ & \textbf{0.7924}$^\dag$ & \textbf{0.4097}$^\dag$ \\
    \midrule
    only brand & 0.5977& 0.7710 & 0.3859 \\
    only color & 0.5867& 0.7626 & 0.3773 \\
    only cate1-4 & 0.6212$^\dag$ & 0.7893$^\dag$ & 0.4050$^\dag$ \\
    brand+color+cate1 & 0.6192$^\dag$ & 0.7863$^\dag$ & 0.4040$^\dag$ \\
    brand+color+cate1-2 & 0.6199$^\dag$ & 0.7898$^\dag$ & 0.4073$^\dag$ \\
    brand+color+cate1-3 & 0.6223$^\dag$ & 0.7910$^\dag$ & 0.4092$^\dag$ \\
    \midrule
    \our$^{-\mathcal{L}_{c2a}}$ & 0.6211$^\dag$ & 0.7882$^\dag$ & 0.4068$^\dag$ \\
    \our$^{-\mathcal{L}_{a2c}}$ & 0.6127$^\dag$ & 0.7846$^\dag$ & 0.3997$^\dag$ \\
    \our$^{-\mathcal{L}_{mlm}}$ & 0.6145$^\dag$ & 0.7851$^\dag$ & 0.4013$^\dag$ \\
    \midrule
    BIBERT$^{+AGREE}$ & 0.6246$^\dag$ & 0.7913$^\dag$ & 0.4112$^\dag$ \\
    \our$^{+AGREE}$ & \textbf{0.6393}$^\dag$ & \textbf{0.8019}$^\dag$ & \textbf{0.4257}$^\dag$ \\
    \bottomrule
  \end{tabular}
\vspace*{-1mm}
\end{table}

\begin{table}
\small
\renewcommand{\arraystretch}{0.4}
\setlength{\abovecaptionskip}{0pt}
  \setlength\tabcolsep{12pt} 
  \caption{Ablation study of query/item aspects on \alipay. $\dag$ indicates significant improvements over BIBERT.}
  \label{tbl:qd}
  \begin{tabular}{lllll}
    \toprule
     & r@100 & r@500 & ndcg@50 \\
    \midrule
    BIBERT & 0.4464 & 0.6284 & 0.2033 \\
    \our & \textbf{0.4667}$^\dag$ & \textbf{0.6592}$^\dag$ & 0.2113$^\dag$ \\
    \midrule
    \our$^{only \  d}$ & 0.4563$^\dag$ & 0.6437$^\dag$ & 0.2105$^\dag$    \\
    \our$^{only \  q}$ & 0.4526 & 0.6366$^\dag$ &  0.2059  \\
    \bottomrule
  \end{tabular}
\end{table}

\noindent%
\textbf{Ablation Study of Loss Function.} We remove each of the three losses from the overall loss to see how important it is. 
In Table \ref{tbl:ab-loss}, we find that: 
(1) The bidirectional prediction losses are beneficial to \our, and excluding either leads to a performance drop. 
(2) The Aspect-to-Content(a2c) prediction is the most helpful, indicating that using aspects as context for content MLM prediction is a feasible way to infuse the aspect information into an item.  
(3) The performance also drops a lot when the vanilla MLM loss is eliminated, indicating the original content semantics without being affected by external information are also important. 

\noindent%
\textbf{Combination with Advanced Fine-tuning Techniques.} 
AGREE \cite{agree} is a recently proposed fine-tuning method that incorporates a local aspect-query matching loss with the original global query-item matching loss. AGREE has not studied how to utilize query aspects, which suits MA-Amazon well since it does not have query aspects. 
Since AGREE concatenates the item aspects with content, it is easy to integrate AGREE during fine-tuning after pre-training with \our. 
The last block in Table \ref{tbl:ab-loss} shows the performance of AGREE alone and combining both. It shows that based on better fine-tuning techniques, \our can achieve better performance. Notably, combining AGREE with methods that conduct aspect classification will not necessarily lead to better performance (Check MTBERT-C and MADRAL-C in Table \ref{main_exp}).

\noindent%
\textbf{Ablation Study of Query/Item Aspects.}
We examine the influence of the query and item aspects in Table \ref{tbl:qd}. 
It shows that both query aspects and item aspects contribute to retrieval performance and the item aspects are more important. Since we only use item aspects during relevance matching, query aspects only take effect during pre-training and could have fewer contributions. 

\vspace*{-3mm}
\section{Conclusion}
In this paper, we propose an effective pre-training method that uses aspects as text strings and conducts mutual prediction between the aspect and content text for multi-aspect retrieval.
In contrast to previous approaches that treat aspect values as categorical IDs, \our can capture the semantic relation between aspects by their text strings and perform finer-grained interactions between item aspect and content by mutual prediction. 
Our experiments on two real-world datasets show that \our can outperform multiple competitive baselines significantly. 
Moreover, we release our enriched Multi-aspect Amazon Product Search dataset to encourage research on multi-aspect dense retrieval.
\vspace*{-2mm}
\begin{acks}
This work was funded by the National Natural Science Foundation of China (NSFC) under Grants No. 61902381, the Youth Innovation Promotion Association CAS under Grants No. 2021100, the project under Grants No. JCKY2022130C039 and 2021QY1701, the Lenovo-CAS Joint Lab Youth Scientist Project. This work was also supported by Ant Group through Ant Innovative Research Program.
\end{acks}